\documentclass[twocolumn,10pt]{tsfp}
\usepackage{flushend}
\usepackage{graphicx}
\usepackage[authoryear,round]{natbib}
\usepackage{fancyhdr}
\usepackage{xcolor}
\usepackage{amsmath}
\usepackage{amssymb}
\usepackage{booktabs}
\usepackage{enumitem}
\usepackage{soul}
\usepackage[normalem]{ulem}

\definecolor{OliveGreen}{RGB}{0,200,0}
\definecolor{DarkBlue}{RGB}{0,0,200}

\setlist[itemize]{label=\textbullet}

\newcommand{\brac}[1]{\langle #1 \rangle}

\newcommand{\biggbrac}[1]{\biggl \langle #1 \biggr \rangle}

\newcommand{\p}{\partial }

\title{Measurements of non-linear energy transfer in canonical and drag-reduced turbulent boundary layers}

\author{Max W. Knoop$^{1}$, Bas W. van Oudheusden$^{1}$, and Rahul Deshpande$^{2,3}$ 
    \affiliation{$^{1}$Faculty of Aerospace Engineering, Delft University of Technology,
	Delft, 2629 HS, The Netherlands\\
    $^{2}$Department of Mechanical Engineering, The University of Melbourne, Parkville 3010, Australia}\\
    $^{3}$School of Engineering, RMIT University, Melbourne 3000, Australia}

\begin{document}

\maketitle   
\thispagestyle{fancy}

\fontsize{9}{11}\selectfont

\section*{ABSTRACT}
Three-dimensional particle-tracking velocimetry (3D-PTV) measurements were used to compute the spectral transport of the Reynolds-stress tensor. 
The experimental framework is validated for a zero-pressure-gradient (ZPG) turbulent boundary layer (TBL) at a friction Reynolds number $Re_\tau = 1020$, demonstrating that the dominant non-linear energy transfer mechanisms are adequately resolved to draw flow physics-based conclusions. 
For the streamwise Reynolds stress in the ZPG TBL, a component-wise decomposition of the non-linear transport term is considered for the first time, which reveals distinct energy transfer mechanisms associated with the spanwise and wall-normal advection.
The same experimental framework was applied to a drag-reduced ($\approx 38\%$) TBL flow, achieved by imposing a steady streamwise-alternating spanwise wall velocity.
This wall forcing causes a strong attenuation of non-linear energy transfer and its shift away from the wall. 
The energy transfer mechanisms remain qualitatively similar to those of the canonical ZPG TBL, suggesting that the existing mechanisms
simply readjust to their new low-turbulent-energy state. 

\begin{figure*}[t]
	\centering
    \includegraphics[width = 1.0\textwidth]{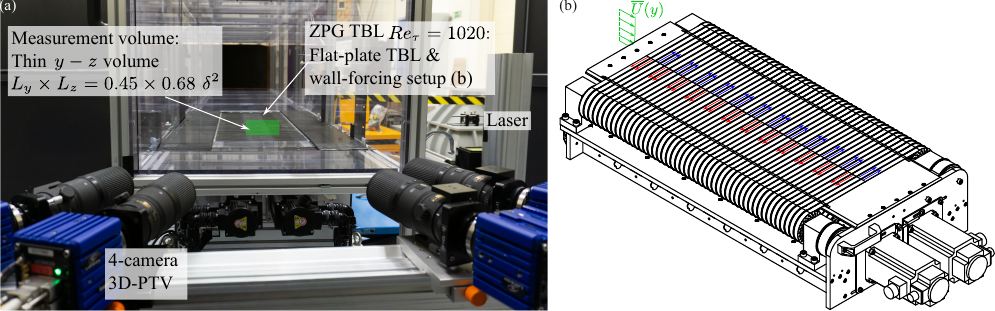}
	\caption{(a) Photograph of the 3D-PTV experimental setup, with the DR wall-forcing setup installed.
    (b) Schematic of the wall-forcing setup deployed \citep{knoop2025response} to impose a steady streamwise-alternating square-wave of steady spanwise wall-velocity. }
	\label{fig:setup}
\end{figure*}
\begin{table*}[t]
\label{Table1}
\centering
{\fontsize{9pt}{9pt}\selectfont
\caption{Overview of the ZPG TBL cases. The DNS dataset is from \citet{sillero2014two} where $dy$ is the minimum grid spacing in $y$.
For 3D-PTV, spatial averaging occurs across the 3D binning volumes of size $(dx)^3$; the filtered DNS is boxcar-filtered to match the 3D-PTV spatial resolution.}
\begin{tabular}{llllllllll}

\hline
\hline 
& \multicolumn{4}{c}{Conditions} & \multicolumn{2}{c}{Domain size} & \multicolumn{2}{c}{Spatial averaging} \\
\cmidrule(r){2-5} \cmidrule(lr){6-7} \cmidrule(l){8-9}
Case & $U_\infty$& $U_\tau$& $\delta$& $Re_\tau$ & $L_x\times L_y \times L_z$  &$L_x \times L_y \times L_z$ & $dx$ &  $dx^+ \times dy^+ \times dz^+ $ \\ 
 & (m/s) & (m/s) & (mm) &  & (mm$^3$) & ($\delta^3$) & (mm) & \\
\hline
Flat-plate TBL (3D-PTV) &  5 & 0.202 & 76 & 1020 & $0.8\times 35 \times 53$ & $0.01\times 0.45 \times 0.68$ & 0.74 & $9.7 \times 9.7 \times 9.7$  \\ 
DNS \citep{sillero2014two}& -- & -- & -- & 1280 & -- & $0.04\times 1.2 \times 2$ & -- & $7.1 \times 0.3 \times 4.4$ \\ 
Filtered DNS & -- & -- & -- & 1280 & -- & $0.04\times 1.2 \times 0.68$ & -- & $9.7 \times 9.7 \times 9.7$ \\ 
\hline
\hline
\end{tabular}
}
\end{table*}

\section*{INTRODUCTION}

The mechanism of turbulent skin-friction drag reduction (DR) via transverse wall forcing has been primarily investigated at low Reynolds numbers \citep{ricco_review_2021}, $Re_\tau \equiv U_\tau \delta/\nu= \mathcal{O}(200-1000)$ and becomes increasingly complex as the scale-separation (i.e. $Re_\tau$) increases. 
Here, $\delta$ is the boundary layer thickness, $U_\tau$ the skin-friction velocity and $\nu$ the kinematic viscosity; viscous scaling using $U_\tau$ and $\nu$ is denoted by the `+' superscript.
For a wide range of wall-forcing conditions, including frequencies/wavenumber away from the DR optimum, a broadband spectral energy attenuation occurs, which broadens further as $Re_\tau$ increases.
The broadband nature of the turbulence spectrum is caused by non-linear momentum (and energy) transfer, and based on high Reynolds number ($Re_\tau = 9700$) experiments, \citet{deshpande2023relationship} hypothesised that this broadband attenuation is associated with an enhanced inter-scale coupling within the actuated TBL.

In this work, we investigate the underlying non-linear energy transfer mechanisms for canonical and drag-reduced TBLs using the spectral energy transport equation.
The spectral turbulent kinetic energy (sTKE) equation \citep{pope2000turbulent} has been widely considered to study wall-bounded turbulence \citep[e.g.,][]{hartel1994subgrid, piomelli1991subgrid, mizuno2016spectra, cho2018scale}, but originates from homogeneous-isotropic turbulence research.
The turbulent boundary layer, however, is highly anisotropic, which motivates our investigation of the spectral transport for the Reynolds stresses that has received less attention \citep{kawata2018inverse, lee2019spectral, chan2021interscale}.
For a given Reynolds stress $\langle u_i u_j\rangle$, the spectral transport equation reads,
\begin{equation}
    \label{eq:spectTransEq}
    \frac{\mathrm{D}\widehat E_{ij}}{\mathrm{D}t} = \widehat P_{ij}   + \widehat T_{ij}  + \widehat \Pi_{ij}  + \widehat D^\nu_{ij} + \widehat \varepsilon_{ij},    
\end{equation}
where $\widehat \cdot$ denotes the spanwise Fourier transform with wavenumber/wavelength $k_z = 2\pi/\lambda_x$, $\widehat E_{ij}$ is the co-spectral density of the Reynolds stress. 
Throughout this manuscript, the mean and fluctuating velocity are indicated by $\langle U_i \rangle$ and $u_i$, where angle brackets denote ensemble and spatial averaging along homogeneous directions.
Subscript $i = (1,2,3) \widehat = (x,y,z)$ correspond to the streamwise, wall-normal and spanwise coordinates with velocity components $(u,v,w)$.

From left to right, the terms in \eqref{eq:spectTransEq} represent: production, turbulent transport, pressure work, viscous diffusion, and dissipation, and these terms are equivalent to the integral Reynolds-stress budget when integrated across wavenumber, e.g. $P_{ij} = \int_{-\infty}^\infty \widehat P_{ij}\mathrm{d}k_z$. 

$\widehat T_{ij}$ is the only non-linear term in \eqref{eq:spectTransEq} and is solely responsible for inter-scale energy transfer; it reads,
\begin{equation}
\label{eq:spectralT}
\widehat T_{ij}(y, k_z) = \mathrm{Re} \biggbrac{ - \widehat u_i^* \frac{\p \widehat{u_ju}_k}{\p \widehat x_k} - \widehat u_j^* \frac{\p  \widehat{u_iu}_k}{\p \widehat x_k}},
\end{equation}
where $\mathrm{Re}$ and $^*$ denote the real component and complex conjugate. 
The (spectral) turbulent transport term arises from the advective non-linearity in the Navier-Stokes equations and comprises both spatial and inter-scale transport. 
Since inter-scale transfer has a net-zero contribution to the integral budget, when $\widehat T_{ij}$ is integrated across all wavenumbers, the spatial transport term $T_{ij}$ in the budget equation is recovered.

For wall-bounded flows, experimental attempts to measure $\widehat T_{ij}$ have been limited to planar 2- or 3-velocity-component particle image velocimetry (2D-2/3C PIV) \citep[e.g.][]{wang2021energy}, or surrogate measures based on single-point time-series \citep[e.g. $\overline{uuu}$ bi-spectra;][]{byers2025identification}.
Such measurements can only be used to compute a subset of the full tensor in \eqref{eq:spectralT}. 
To the best of our knowledge, experimentally, the complete $\widehat T_{ij}$ term for the Reynolds stresses has been successfully computed only by \citet{kawata2018inverse}, using 2D-3C PIV and by scale-decomposing the integral budget.
Owing to the (spectral) gradients in all directions, directly computing \eqref{eq:spectralT} requires instantaneous, well-resolved, and low-noise volumetric (3D) flow-field measurements.
To satisfy these requirements, we conducted 3D particle tracking velocimetry (3D-PTV) in a zero-pressure gradient (ZPG) TBL at $Re_\tau \approx 1000$. 

The aim of this work is twofold. 
The first part establishes our experimental framework to resolve all the terms in \eqref{eq:spectTransEq} except pressure transport, with particular focus on $\widehat T_{ij}$. 
Our framework is validated at low $ Re_\tau$ using smooth-wall ZPG TBL measurements, which are compared with direct numerical simulation (DNS).
In the second part, this framework is applied to evaluate the impact of DR induced by wall forcing on the non-linear energy transfer mechanisms.

\section*{VOLUMETRIC FLOW-FIELD MEASUREMENTS AND NUMERICAL DATASETS}
Experiments were conducted in the W-tunnel at the Delft University of Technology \citep[details in][]{knoop2025response} at a free-stream velocity of $U_\infty = 5$\:m/s, achieving a ZPG TBL of $Re_\tau \approx 1000$ at the measurement station located 3.05\:m downstream of the boundary-layer trip (P40 sandpaper).
Three cases were considered: a flat-plate TBL flow and a TBL flow over the non-actuated and actuated wall-forcing setup \citep{knoop2025response}, where the forcing effect was fully established.
Figure~\ref{fig:setup}(a) shows a photograph of the experimental setup with the wall-forcing mechanism installed flush with the tunnel floor, while Figure~\ref{fig:setup}(b) depicts a schematic of the forcing setup (further details provided in the results).

Two-pulse particle tracking velocimetry (PTV) \citep{novara2023two} was used to obtain three-dimensional three-component (3D-3C) flow-field measurements in a thin measurement volume (essentially a \emph{thick laser-sheet}).
The measurement volume was oriented in the cross-stream $(y,z)$ plane, and had dimensions $L_x\times L_y \times L_z = 3\times 35 \times 55$\:mm$^3$ (present analysis only considered a 0.8\:mm region of the full $L_x$). 
Four sCMOS cameras ($2560\times 2160$ px$^2$; 6\:\textmu m; 16-bit), mounted with Scheimpflug adapters and 200-mm objectives at f/11, were placed in a linear configuration, rotated by $\pm 30^\circ,15^\circ$ degrees around the $y$-axis, for a total system aperture of $60^\circ$.
Illumination of the water-glycol tracer particles was provided by a 200\:mJ/pulse Nd:YAG laser. 
The low-light Gaussian tails of the laser beam were removed using a knife-edge filter. 
To mitigate low illumination for the two cameras in backscatter, the sheet was reflected using a mirror to provide double-pass illumination.
A total of 1000 uncorrelated dual-pulse snapshots were acquired at 8\:Hz (separation corresponds to 9 TBL turnover times), with a time separation of $90$\:\textmu s between image pairs (20 pixel displacements in the freestream).
Each snapshot contained approximately $100\:000$ tracked particles. 
To obtain gridded data, first-order polynomial binning was used with $40^3$ voxel Gaussian-weighted bins with a 75\% overlap factor. 
The viscous-scaled bin size was $dx^+ = 9.7$.

\begin{figure*}[t]
    \centering
    \includegraphics[]{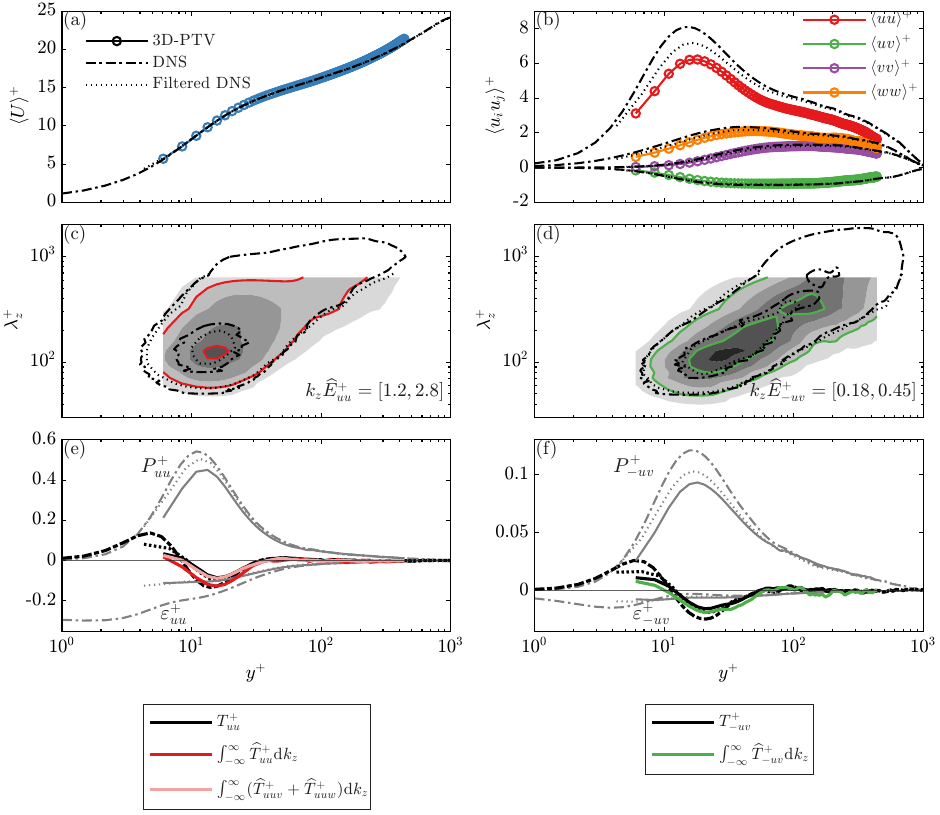}
    \caption{Validation of the flat-plate TBL with (filtered) DNS \citep{sillero2014two}. Profiles of (a) mean streamwise velocity and (b) Reynolds stresses; (c,d) premultiplied energy spectra of $\brac{uu}$ and -$\brac{uv}$, grey filled and coloured line contours correspond to 3D-PTV; (e,f) integral budget profiles of production and dissipation (grey) and turbulent transport matching the linestyles/colours in (a-d). In (e) $\widehat T_{uu} = \widehat T_{uuu} +\widehat T_{uuv} + \widehat T_{uuw}$.}
    \label{fig:validation}
\end{figure*}

Table\:1 provides an overview of the ZPG TBL cases used to validate our experimental dataset and the computation of \eqref{eq:spectTransEq}.
The same measurement setup and data processing procedure were also employed to acquire 3D-3C measurements over the spanwise wall forcing setup (for both non-actuated and actuated cases).
First, we compare the flat-plate TBL (3D-PTV) experiment with direct numerical simulations (DNS) by \citet{sillero2014two} of a ZPG TBL at comparable $Re_\tau = 1280$ for validation of the measurements and analysis framework. 
To analyse the DNS dataset, thin volumes similar to the measurement volume ($L_x \times L_y \times L_z = 0.04\delta \times 1\delta \times 2\delta$) were extracted at five equally-spaced streamwise locations between $x/\delta = 0.73- 1.57$ (minimal variation of $Re_\tau = 1275-1290$) for thirteen instantaneous velocity snapshots of uncorrelated time-steps.
To verify the effect of spatial-averaging of the volumetric binning, a filtered DNS dataset was obtained by interpolating on grid spacing ($\Delta x = dx/4$) and averaging using a 3D boxcar filter to match the $dx^+ = 9.7$ spatial resolution of the 3D-PTV dataset. 
The spanwise domain size, $L_z$, was also matched to the experiment.

\section*{VALIDATION OF THE EXPERIMENTS}
1st- and 2nd-order turbulence statistics are shown in Figures~\ref{fig:validation}(a,b).
The $\langle U\rangle$ profile in Figure~\ref{fig:validation}(a) is well resolved down to $y^+ = 6$ by 3D-PTV.
The filtered DNS in Figure~\ref{fig:validation}(b) reveals the effect of spatial averaging.
Namely, a reduction of the $\langle uu\rangle$ inner-peak, while the effect on the other Reynolds stresses remains small.
Compared to the filtered DNS, 3D-PTV shows greater attenuation of the inner peak, which may be attributed to the increased spatial averaging volume in Gaussian-weighted binning, extending the effective bin size by approximately a factor 2. 
In the outer-layer, the overall lower $\langle uu\rangle$ is mainly due to the lower $Re_\tau$ of the experimental dataset (see table~\ref{Table1}). 
Qualitatively, the same behaviour is apparent in the premultiplied energy spectra $k_zE_{uu}$ and $k_zE_{-uv}$ in Figures~\ref{fig:validation}(c,d), showing only minor deviations from the DNS.

Figures~\ref{fig:validation}(e,f) show wall-normal profiles of the integral budget for $\brac{uu}$ and $-\brac{uv}$. 
The production, dissipation and turbulent transport are computed as 
\begin{equation}
    \label{eq:integralBudget}
    \begin{aligned}
    & P_{ij} = \langle u_i u_k\rangle \frac{\p \langle U_j\rangle}{\p x_k} + \langle u_j u_k\rangle \frac{\p \langle U_i\rangle}{\p x_k},\\
    & \varepsilon_{ij} = -2\nu \biggbrac{\frac{\p u_i}{\p x_k} \frac{\p u_i}{\p x_k}},\\
    & T_{ij} = \frac{\p\brac{u_i u_j u_k}}{\p x_k},
    \end{aligned}
\end{equation}
all of which are plotted in grey shading in Figure \ref{fig:validation}.
For the 3D-PTV experiment, the integral budget terms are attenuated by spatial averaging, most clearly for production (in grey), while closely matching the filtered DNS for $\varepsilon_{ij}$ (in grey) and $T_{ij}$ (in black).

\begin{figure*}
    \centering
    \includegraphics[width = \textwidth]{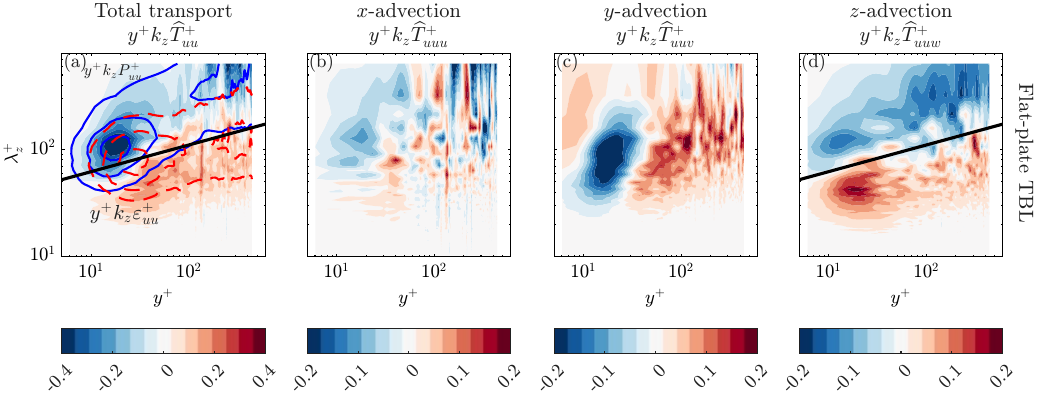}
    \caption{Premultiplied non-linear spectral transport spectra of $\brac{uu}$ for the flat-plate TBL. (a) Total transport $\widehat T_{uu}$ and (b-d) component-wise decomposition associated with advection in each coordinate direction, i.e $\widehat T_{uuu}$, $\widehat T_{uuv}$, $\widehat T_{uuw}$. Overlayed line spectra in (a) of production (blue solid) with levels $y^+ k_z\widehat P_{uu}^+ = [0.2, 0.5, 0.8]$ and dissipation (red dashed) with levels $y^+ k_z\widehat \varepsilon_{uu}^+ = [-0.09, -0.13, -0.17]$. Black lines in (a,d) indicate dissipation scaling $\lambda_z \propto y^{1/4}$.}
    \label{fig:componentwiseTransport}
\end{figure*}

Recall that the integration of the Fourier decomposed term of nonlinear transport, \emph{i.e.} $T_{ij} = \int_{-\infty}^\infty\widehat T_{ij}\mathrm{d}k_z$, should ideally recover the integral term. 
The coloured lines in Figures~\ref{fig:validation}(e,f) show the integral of $\widehat T_{ij}$ to validate its computation.
In Figure~\ref{fig:validation}(e), the integral of $\widehat T_{uu}$ is enhanced with respect to the direct calculation of $T_{uu}$.
We later show (refer to Figure~\ref{fig:componentwiseTransport}) that the effect of the $\p/\p x$ gradient (i.e., $k=1$) is small, and the enhancement is likely due to low signal-to-noise ratio for this component, increasing the sensitivity to noise in the spectral and gradient terms.
To confirm this effect, the light red line excludes the streamwise gradients by only adding the $k=2,3$ terms, i.e. $\widehat T_{uuv} +\widehat T_{uuw}$.
This result now closely matches the direct computation of $T_{uu}$.  
The integral of $\widehat T_{-uv}$ in Figure~\ref{fig:validation}(f) shows better agreement with direct budget calculation; we also verified that $\widehat T_{-uvu}$ does have a contribution to $\widehat T_{-uv}$.
We conclude from this validation exercise that, although the experimental reconstruction of the \emph{absolute} magnitudes of these higher-order statistics (especially the spectral gradients of velocity correlations) remains challenging, their qualitative behaviour is well captured, providing confidence in our 3D-3C measurements.

\section*{COMPONENT-WISE ENERGY TRANSFER}

We consider the spectral energy transfer for the streamwise normal stress ($\langle uu \rangle$), which is characteristic of the near-wall streaks and energetic near-wall cycle.  
Figure~\ref{fig:componentwiseTransport}(a) shows the premultiplied $\widehat T_{uu}$ in coloured contours, while the line-contours show production (blue solid) and dissipation (red dashed).  
Figure~\ref{fig:componentwiseTransport}(a) is consistent with the well-known \citeauthor{richardson1922weather}-\citeauthor{kolmogorov1941local}  forward energy cascade from large to small scales. 
Energy is produced (blue line-contours) at larger scales at the top-side of the graph, with the near-wall production peak at $\lambda_z^+ \approx 100$ around $y^+ = 15$ associated with the energetic near-wall cycle, while energy is dissipated (red dashed-line-contours) at smaller scales at the bottom-side of the graph.
$\widehat T_{uu}$ drives the inter-scale transfer from the $\widehat P_{uu}$ dominated scales, where $\widehat T_{uu}<0$ sinks energy, towards the $\widehat \varepsilon_{uu}$ scales, where $\widehat T_{uu}>0$ gains energy, i.e. energy transfers from top to bottom along the vertical $\lambda_z$-axis.

$\widehat T_{ij}$ derives from the non-linear advection term in the Navier-Stokes equation. Owing to this non-linearity, underlying $\widehat T_{ij}$ are individual scale-interactions between three triadically coupled wavenumbers. 
While the discussion and derivation of that formulation is beyond the scope of these proceedings, we briefly consider the physical interpretation of the energy-transfer process as an advective interaction between two modes \citep{young2024inter,ding2025mode}.
A scale-interaction occurs between $u_j$ at one scale that is advected by the $u_k$ (note the $\p /\p x_k$ gradient in \eqref{eq:spectralT}) at a second scale to cause energy transfer to a third $u_i$ mode. 
$\widehat T_{ij}$ captures the net-energy transfer to scale $k_z$ by integrating across all these underlying interactions. 
Omitting the summation over $k$ gives $\widehat T_{ijk}$, which allows us to isolate the effects of advection in the separate coordinate directions. 
To the best of our knowledge, such a decomposition has not been considered previously.

Figures~\ref{fig:componentwiseTransport}(b-d) show the component-wise energy transfers, i.e. $\widehat T_{uuu}$, $\widehat T_{uuv}$, and $\widehat T_{uuw}$, which reveal completely different energy transfer mechanisms for advection in each direction.
Whereas streamwise advection ($\widehat T_{uuu}$) has a near-zero contribution, the cross-stream components $\widehat T_{uuv}$ and $\widehat T_{uuw}$ dominate the non-linear energy transfer process.
Energy transfer by spanwise advection ($\widehat T_{uuw}$), shown in Figure~\ref{fig:componentwiseTransport}(d), follows a dissipative scaling ($\lambda_x \propto y^{1/4}$ in the solid black line), transferring from inertial scales (loss in blue) to small scales (gain in red), and is therefore governing the classic \citeauthor{richardson1922weather}-\citeauthor{kolmogorov1941local} forward energy cascade.
Wall-normal advection $\widehat T_{uuv}$ in Figure~\ref{fig:componentwiseTransport}(d) is characterised by a strong energy loss around $y^+ = 15$ for $50 \lesssim \lambda_z^+ \lesssim 200$, flanked by a gain above and below the near-wall peak, reflecting energy transfer away from the energetic streaks and near-wall cycle.
Consequently, wall-normal advection underlies two energy transfer pathways.
One towards the log-layer for $y^+ >30$ with a broadband energy gain ($\lambda_z^+\gtrsim 20$) and no clear forward/inverse energy cascade.
The second pathway is towards the near-wall region $y^+ < 10$ at significantly larger scales($\lambda_z^+\gtrsim 100$), i.e. an inverse energy cascade.
Interestingly, a pure inverse energy cascade reported by \citet{chan2021interscale} for the Reynolds shear stress $-\brac{uv}$, albeit limited to the $\widehat T_{-uv}$ and not its three components.
Our results, therefore, suggest that wall-normal advection plays a role in the inverse energy-transfer mechanisms.


\section*{DRAG-REDUCED ENERGY TRANSFER MECHANISMS}

\begin{figure*}
    \centering
    \includegraphics[width = \textwidth]{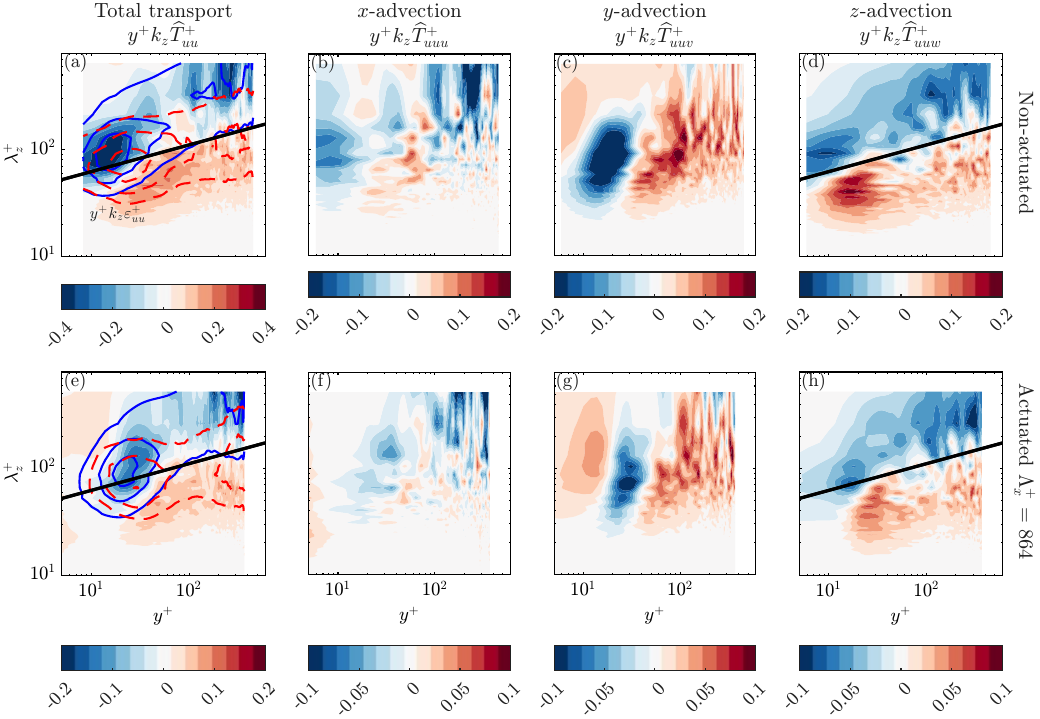}
    \caption{Premultiplied non-linear spectral transport spectra of $\brac{uu}$ for the wall-forcing cases. Top row (a-d) non-actuated and bottom row (e-h) actuated cases. (a-d) Layout for matches the flat-plate TBL in Figure~\ref{fig:componentwiseTransport}, while for  (e-h) colour and contour levels are 50\% of the non-actuated case.}
    \label{fig:DRcomponentwiseTransport}
\end{figure*}

The effect of turbulent drag reduction (DR) by spanwise wall forcing is considered next.
Wall-forcing was implemented using the setup reported in \citet{knoop2025response}, which is schematically shown in Figure~\ref{fig:setup}(b).
The setup consists of a series of belts, rotating in alternating positive/negative spanwise direction, to impose a spatially varying square-wave of spanwise wall-velocity, analogous to the sinusoidal form typically studied $W_w = A\sin(\mathrm{k}_x x)$.
$A$ and $\Lambda_x = 2\pi/\mathrm{k}_x$ are spanwise velocity amplitude and forcing wavelength.
\citet{knoop2025response} confirmed that the forcing effect, regarding the $\Lambda_x$ effect on turbulence statistics and DR, aligns with the literature on sinusoidal forcing \citep{Viotti2009StreamwiseReduction}.
For the actuated case, a near-optimal DR case \citep[$\Lambda_x \approx 1000$;][]{Viotti2009StreamwiseReduction} was selected with $\Lambda_x^+ = 864$ and $A^+ = 12.8$ to achieve a strong control effect; for this case \citet{knoop2025response} reported a DR of approximately 38\%. 
A non-actuated case (i.e., no belt rotation, so $A=0$) was measured to serve as a reference. 
To bring out absolute changes of the wall-forcing, a viscous scaling based on the reference $U_{\tau 0}$ of the non-actuated case is adopted.

\begin{figure}
    \centering
    \includegraphics[width=\linewidth]{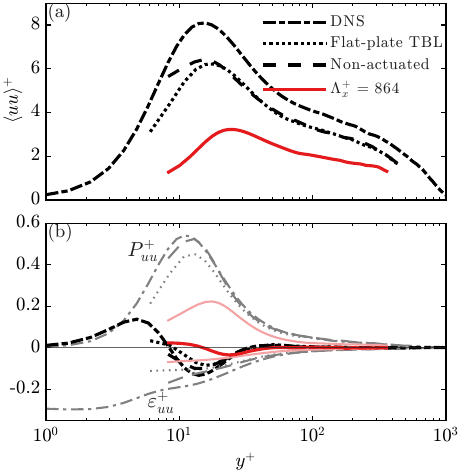}
    \caption{Turbulence statistics comparing the wall-forcing (non-actuated and actuated) cases to the reference ZPG TBLs. (a) Streamwise Reynolds stress, and (b) integral budgets. (Light) coloured lines in (b) indicate the actuated case.} 
    \label{fig:statisticsDR}
\end{figure}

The TBL over the non-actuated setup is first compared against the (smooth) flat-plate TBL to qualitatively establish the negligible influence of the actuation surface (\emph{i.e.}, tunnel wall versus belts of the forcing setup).
Figure~\ref{fig:statisticsDR} compares $\brac{uu}$ and its integral budget for the different cases.
With respect to the flat-plate TBL, a slight enhancement of the magnitudes near the wall ($y^+ \lesssim 10$) occurs, which is due to the flow over the complex actuation surface; details are discussed in \citet{knoop2025response}. 
Given these small differences, the actuation effect is assessed by comparing the actuated case to the non-actuated case. 
The non-linear transport for the non-actuated case in Figure~\ref{fig:DRcomponentwiseTransport}(top row) is consistent with that of the flat-plate TBL in Figure~\ref{fig:componentwiseTransport}, which confirms that the existing mechanisms are not significantly altered by the actuation surface geometry.

Figure~\ref{fig:statisticsDR} clearly shows the effect of wall-forcing on the integral turbulence statistics.
The inner-peak in Figure~\ref{fig:statisticsDR}(a) is reduced by $\approx 50\%$ due to the dampening of the near-wall cycle, and its outward shift from $y^+ =15$ to 25 indicates a thickening of the viscous sublayer; these effects agree with the established literature \citep{ricco_review_2021}.
Similar effects are seen for the integral budget in Figure~\ref{fig:statisticsDR}(b).

The effect of DR on the non-linear energy transfer is shown in Figure~\ref{fig:DRcomponentwiseTransport}.
To provide a fair comparison in view of the strongly reduced statistics, in Figure~\ref{fig:DRcomponentwiseTransport}, the colorscale and contour levels used for the actuated case (bottom row) are half (i.e. 50\%) of those for the non-actuated case (top row).
For the actuated case, $\widehat T_{uu}$ and its components are reduced by more than half, and the transport mechanisms shift away from the wall in view of the thickening of the viscous sublayer.  
Unlike the non-actuated case, a gain of $\widehat T_{uuu}$ occurs in the near-wall region for $y^+ < 20$. 
This effect can be attributed to the outward movement of the near-wall region of positive transport, i.e. $T_{uu}>0$ where $y^+ < 10$ (not captured but seen for DNS) in Figure~\ref{fig:statisticsDR}(b).
Although the energy transfer is significantly reduced, its qualitative trends remain the same in the actuated case.
Our results qualitatively suggest that, while the near-wall cycle is strongly attenuated by the forcing, broadly similar energetic dynamics remain and the TBL adjusts to the new low-energy state. 
In this process, the energy transfer mechanisms may not be altered by the wall-forcing.

\section*{CONCLUSIONS}
An experimental framework was provided to compute the spectral transport of the Reynolds stresses using 3D-PTV. 
The measurements were validated using low-$Re_\tau$ DNS, and despite spatial averaging due to the finite-sized 3D bins, the statistics and spectral budget terms are well resolved.   
This framework can subsequently be applied to cases where DNS are computationally intensive, e.g., high-$Re_\tau$ ZPG TBLs or non-canonical cases such as roughness or flow control.

Our analysis of the component-wise energy transfer of the streamwise normal stress reveals the importance of identifying the mechanisms associated with advection by three separate velocity components. 
Further elucidating these distinct mechanisms is deemed valuable, for example, focusing on why spanwise advection (i.e., $\widehat T_{uuw}$) drives the classic forward energy cascade. 
Wall-normal advection ($\widehat T_{uuv}$) causes energy transfer from the energetic $\brac{uu}$-peak to the log-layer ($y^+>30$) and an inverse energy cascade towards the near-wall region ($y^+<10$).
This second pathway suggests the importance of wall-normal advection in the underlying inverse energy-transfer mechanisms.


In the drag-reduced TBL, while the magnitude of the energy transfer is strongly attenuated, the qualitative trends remain unchanged. 
These results suggest that the energy-transfer mechanisms are not altered by the wall-forcing but simply readjust to the low-energy drag-reduced state.
The broadband energy attenuation discussed by \citet{deshpande2023relationship} may likely be explained by the strong $\widehat T_{uu}$ reduction induced by the wall motion.

A firm conclusion, however, can only be drawn after analysis of the underlying triadic interactions, which we intend to report in a full manuscript in the near future.
Similar to the flat-plate ZPG, investigating the role of wall-normal advection and, owing to its direct link with skin friction, extending the analysis to Reynolds shear stress are other potential research directions we intend to pursue.

\bibliographystyle{tsfp}
\bibliography{tsfp}

\newpage

\end{document}